\documentclass[letterpaper,twocolumn,10pt]{article}
\usepackage{float}
\usepackage{usenix-2020-09}

\usepackage{amsmath}
\usepackage{array}
\usepackage{booktabs}
\usepackage{graphicx}
\usepackage{tabularx}
\usepackage{balance}
\usepackage{fontawesome5}
\usepackage{multirow}
\usepackage{tikz}
\usepackage{url}
\usetikzlibrary{arrows.meta,positioning}

\newcommand{\tool}{\textsc{WGPULens}}

\begin{document}

\date{}


\title{\Large \bf What Browsers Do in the Shaders: A Measurement Study of WebGPU Privacy}

\author{
{\rm Igor Santos-Grueiro}\\ International University of La Rioja
}

\maketitle

\begin{abstract}
WebGPU lets ordinary web pages run GPU workloads through a validated
programming model. Validation protects memory safety, but shared browser,
driver, OS, and GPU state can still expose privacy-relevant signals. We present
\tool{}, a framework for measuring those signals across controlled scenarios,
browser-native co-residency, a participant field study, public page loads, and
mitigation policies. Our framework separates measurements: controlled
scenarios support leakage, boundary, and mitigation claims; participant runs
support deployment, compatibility, and fingerprintability; and a Tranco crawl
measures WebGPU exposure in real-world pages.

Our controlled results identify persistent pipeline compilation state as the
clearest surface. Cold/warm pipeline probes reveal prior compilation state
across selected origin, profile, and browser placements. Controlled
browser/native experiments also show native GPU activity can be inferred from
browser-visible observables under labeled workloads. Other resource probes
provide weaker positive results and negative controls.

The participant field study shows active WebGPU behavior is highly distinctive
within the sample, with deterministic components stable within runs and lower
exact stability across repeated visits. A page-load crawl finds WebGPU use
mainly as adapter probing and static
support code, with no observed page-load shader, pipeline, queue, query, or map
activity. Mitigation pilots identify source-level key separation as a proxy for
evaluating pipeline-cache partitioning. Overall, \tool{} shows that
WebGPU privacy analysis must be surface-specific: browsers need to measure which
GPU state crosses which boundary, which browser-visible signals reveal it, and
what the corresponding mitigations cost.

\end{abstract}

\section{Introduction}
\label{sec:introduction}

WebGPU brings GPU computation to ordinary web pages. A page can allocate
buffers and textures, compile WGSL shaders, create render and compute
pipelines, and submit GPU work through a standard queue. This supports local
ML inference, media processing, visualization, games, simulation, and
editing~\cite{mdn-webgpu,webdev-webgpu-major-browsers,webllm,llamaweb}.

Canvas and WebGL already showed that graphics APIs expose browser and
hardware variation: rendering hashes reveal graphics-stack differences
~\cite{pixel-perfect}, OS/hardware features enable cross-browser
fingerprints~\cite{cross-browser-fp}, DrawnApart identifies GPUs from WebGL
behavior~\cite{drawnapart}, and renderer strings are privacy-sensitive
~\cite{mdn-webgl-debug-renderer-info,khronos-webgl-debug-renderer-info}.
WebGPU expands this surface from rendering to programmable workloads,
dispatch shape, atomics, pipelines, and queue progress.

WebGPU validation protects memory safety: the API and WGSL validation reject
invalid resource use, out-of-bounds access, uninitialized reads, and undefined
shader behavior~\cite{w3c-webgpu,w3c-wgsl}. That validation still leaves
privacy-relevant signals. We measure what remains observable inside the valid
API boundary: adapter exposure, resource contention, queue progress, completion
behavior, and cold or warm shader/pipeline creation under ordinary JavaScript
and valid WGSL.

Recent work establishes browser-exposed GPUs as a practical privacy surface.
GPU.zip shows cross-origin leakage through graphical compression
~\cite{gpuzip}; WebGPU-SPY and Giner et al. build WebGPU cache attacks
~\cite{webgpu-spy,giner-webgpu-cache}; AtomicIncrement and LockedApart use
WebGPU or compute behavior for device fingerprinting
~\cite{atomicincrement-webgpu,lockedapart}; native GPU work shows timer-free
cache primitives and residual-memory risks~\cite{invalidate-compare,leftoverlocals}.
Browsers and standards bodies need measurements that separate these mechanisms
and name the boundary and observable for each result.

We build \tool{} to collect those measurements. \tool{} generates valid WGSL
probes, runs synthetic web and native victims, coordinates browser boundaries,
records high-resolution, coarse, and ordinal observations, reconstructs
participant uploads, audits page loads, and computes leakage, fingerprinting,
and cost metrics. Controlled scenarios support leakage, boundary,
browser/native, and mitigation claims. The participant field study measures
deployment, heterogeneity, fingerprintability, and repeated-run behavior. A
Tranco top-10k crawl records page-load WebGPU exposure.

Controlled results identify persistent WebGPU pipeline compilation state as
the clearest surface. Windows same-page cold/warm calibration reaches AUROC
0.986 on AMD/RDNA-3 and 0.961 on NVIDIA/Lovelace, with selected web, profile,
Chromium-family, and M1/Metal placements also positive. Browser/native
experiments show synthetic native GPU activity visible through WebGPU under
controlled co-residency, with M1 active/idle reaching macro-F1 0.946 over
coarse-frame observables. Secondary workload and non-pipeline rows are weaker
and more platform-specific.

The participant and crawl measurements complement the controlled results with
evidence from deployed browsers and public pages. Across two participant
campaigns, the 1,095 deduplicated completed records yield 1,095 distinct active
WebGPU signatures. Relative to a browser/OS baseline, these features add 7.136
bits of empirical entropy. Repeated measurements show stable deterministic
components within runs and more limited exact stability across repeated
local-storage identifiers. The Tranco crawl finds 56 WebGPU-positive page-load
records, mostly adapter probes and static support code, with no observed
page-load shader, pipeline, queue, query, or map activity.

We make the following contributions.

\begin{itemize}
  \item \textbf{Evidence design and measurement framework.} \tool{} measures
  static exposure, active contention, and persistent pipeline state with valid
  JavaScript/WGSL probes across controlled, participant, and crawl datasets.

  \item \textbf{Controlled persistent-state and co-residency results.} We show
  pipeline cold/warm separability, selected positive boundary rows, and
  controlled browser/native inference with synthetic OpenCL and Metal workloads.

  \item \textbf{Participant and page-load measurements.} Participant runs show
  active WebGPU fingerprintability within the sample; the Tranco crawl shows
  page-load exposure mostly through adapter probing and static support code.

  \item \textbf{Mitigation case study.} Source-level key separation
  approximates pipeline cache partitioning and measures leakage/cost
  tradeoffs.

  \item \textbf{Artifact availability.} We release an anonymous artifact with
  code, normalized derived data, reproduction scripts, and the extension for
  the full study
  (\url{https://anonymous.4open.science/r/WGPULens/}).
\end{itemize}

\section{Background and Threat Model}
\label{sec:background-threat-model}

WebGPU gives web pages a modern GPU programming model. A page can request a
\texttt{GPUAdapter}, obtain a \texttt{GPUDevice}, create buffers and textures,
compile WGSL shader modules, build compute or render pipelines, record
commands, and submit command buffers to a \texttt{GPUQueue}. This model is now
deployed across major browser families, with backend and platform differences
across Chrome, Edge, Firefox, and Safari~\cite{mdn-webgpu,
webdev-webgpu-major-browsers,chrome-webgpu-release,firefox141-webgpu,
webkit-safari26-webgpu,gpuweb-implementation-status}.

The logical object model abstracts over a shared software and hardware stack.
Chromium-family browsers route WebGPU through Dawn; Firefox routes through
wgpu; Safari/WebKit uses the Apple platform stack. These implementations
translate WebGPU operations to D3D12, Metal, or
Vulkan, then to drivers, OS GPU schedulers, memory managers, and physical
resources shared with other tabs, origins, profiles, browsers, and native
applications.

\subsection{Safety and Observability}

WGSL and WebGPU validation define the safety boundary. Shader code is typed,
resource bindings are explicit, and shader-module or pipeline creation can fail
before execution if validation constraints are violated~\cite{w3c-webgpu,
w3c-wgsl}. Our experiments stay inside this boundary: valid JavaScript and
WGSL, no out-of-bounds accesses, uninitialized reads, undefined shader
behavior, intentional data races, browser exploits, or driver memory disclosure.

The privacy question starts after validation succeeds. A valid page can observe
features and limits, queue progress, completion ordering, coarse frame
behavior, and cold/warm shader or pipeline creation.
\subsection{Observable GPU State}

\tool{} measures static exposure, active contention, and persistent state.
Static exposure includes availability, adapter metadata, features, limits,
formats, shader support, timestamp-query exposure, WebGL capabilities, and
rendering signatures~\cite{mdn-gpuadapterinfo,
khronos-webgl-debug-renderer-info,mdn-webgl-debug-renderer-info}.
Active-contention probes stress storage buffers, textures, workgroup memory, arithmetic
pipelines, dispatch shape, barriers, atomics, and queue progress. Persistent
state focuses on shader and pipeline compilation: pipeline creation may follow
a cold-compilation path or reuse cached state depending on browser, backend,
driver, and cache policy.

The WebGPU specification recognizes timing and workload-identification risks,
including \texttt{timestamp-query} and constrained global resources shared by
programs and pages~\cite{w3c-webgpu}. We therefore measure explicit timing,
coarse browser observables, and ordinal or categorical outcomes.

\subsection{Threat Model}
\label{sec:threat-model}

The attacker controls one web origin and runs ordinary JavaScript and valid
WGSL through WebGPU. The attacker has no native code, browser extension,
privileged API, hardware counters, browser or driver exploit, or access to
victim memory. Victims are controlled workloads placed in the same page,
another origin, an iframe, another tab, a private window, another browser, or a
native GPU application.

The evaluated inference targets are limited to activity, workload family, resource class,
phase, scheduling state, or previous pipeline compilation state under
controlled workloads. Real-user tracking, secret recovery, live-website victim
fingerprinting, pixel stealing, and arbitrary data extraction are outside the
evaluated claims.

\section{\tool{} Methodology and System}
\label{sec:measurement-model}

\tool{} uses a single browser-side measurement core with three measurement
paths. Controlled experiments fix browser, profile, GPU path, origin relation,
victim placement, timer mode, and mitigation condition; hidden labels encode
the boundary, victim state, pipeline state, workload class, and mitigation
condition. Participant runs reuse the same core in an opt-in page to assess
deployment, compatibility, heterogeneity, repeatability, and fingerprintability.
The crawl records page-load WebGPU exposure. Table~\ref{tab:evidence-design}
sets the claim boundary used throughout the paper.

\begin{table}[t!]
\centering
\caption{Evidence design. Each dataset answers a different question.}
\label{tab:evidence-design}
\footnotesize
\setlength{\tabcolsep}{3pt}
\renewcommand{\arraystretch}{1.04}
\begin{tabular}{@{}
  >{\raggedright\arraybackslash}p{0.21\linewidth}
  >{\raggedright\arraybackslash}p{0.35\linewidth}
  >{\raggedright\arraybackslash}p{0.32\linewidth}
  @{}}
\toprule
Dataset & Supports & Does not support \\
\midrule
Controlled scenarios & Leakage, boundary, browser/native, workload, and
mitigation claims under labeled placements & Prevalence in deployed users or
sites \\
Participant field study & Compatibility, browser/GPU heterogeneity,
sample-level fingerprintability, and repeated local-storage identifiers &
Cross-context victim inference or mitigation claims \\
Page-load crawl & WebGPU API exposure, static support code, and source context
on public pages & Leakage, intent, or interactive WebGPU workloads \\
Mitigation pilots & Direction and leakage/cost tradeoffs under matched
controlled rows & Effectiveness of a browser-engine cache-key patch \\
\bottomrule
\end{tabular}
\end{table}

\begin{figure*}[t!]
\centering
\resizebox{0.96\textwidth}{!}{%
\begin{tikzpicture}[
  font=\scriptsize,
  corebox/.style={draw=black!72, line width=0.45pt, rounded corners=4pt,
    fill=black!5, minimum width=8.75cm, minimum height=2.28cm},
  input/.style={draw=black!62, line width=0.34pt, rounded corners=3pt,
    fill=black!1, align=center, inner sep=4pt, text width=2.25cm,
    minimum height=0.82cm},
  module/.style={draw=black!68, line width=0.34pt, rounded corners=3pt,
    fill=white, align=center, inner sep=4pt, text width=2.18cm,
    minimum height=0.88cm},
  stage/.style={draw=black!70, line width=0.38pt, rounded corners=3pt,
    fill=black!3, align=center, inner sep=4pt, text width=1.66cm,
    minimum height=0.94cm},
  result/.style={draw=black!58, line width=0.32pt, rounded corners=3pt,
    fill=black!1, align=center, inner sep=3pt, text width=1.72cm,
    minimum height=0.80cm},
  flow/.style={-{Latex[length=2.2mm,width=1.55mm]}, line width=0.45pt,
    draw=black!75}
]
\node[corebox] (core) at (4.2,0) {};
\node[font=\bfseries\scriptsize] at (4.2,-0.93)
  {\faWindowMaximize\quad Browser measurement core};

\node[module] (meta) at (1.3,-0.08)
  {{\large\faFingerprint}\\[2pt]\textbf{Metadata}};
\node[module] (runner) at (4.2,-0.08)
  {{\large\faMicroscope}\\[2pt]\textbf{Probe runner}};
\node[module] (recorder) at (7.1,-0.08)
  {{\large\faIcon{file-alt}}\\[2pt]\textbf{Event recorder}};

\node[input] (specs) at (1.3,2.02)
  {{\large\faLayerGroup}\\[2pt]Probe specs};
\node[input] (placement) at (4.2,2.02)
  {{\large\faGlobe}\\[2pt]Placement};
\node[input] (victims) at (7.1,2.02)
  {{\large\faDesktop}\\[2pt]Victims};

\node[stage] (trace) at (9.95,-0.08)
  {{\large\faIcon{file-alt}}\\[2pt]\textbf{JSON trace}};
\node[stage] (analyzer) at (12.55,-0.08)
  {{\large\faSearch}\\[2pt]\textbf{Analyzer}};

\node[result] (leakage) at (15.25,1.38)
  {{\normalsize\faFaucet}\\[-1pt]\textbf{Leakage}\\AUROC, F1};
\node[result] (fingerprint) at (15.25,-0.08)
  {{\normalsize\faFingerprint}\\[-1pt]\textbf{Fingerprint}\\entropy};
\node[result] (cost) at (15.25,-1.54)
  {{\normalsize\faStopwatch}\\[-1pt]\textbf{Cost}\\latency};

\draw[flow] (specs.south) -- (meta.north);
\draw[flow] (placement.south) -- (runner.north);
\draw[flow] (victims.south) -- (recorder.north);
\draw[flow] (meta.east) -- (runner.west);
\draw[flow] (runner.east) -- (recorder.west);
\draw[flow] (recorder.east) -- (trace.west);
\draw[flow] (trace.east) -- (analyzer.west);
\draw[flow] ([yshift=0.32cm]analyzer.east) -- (leakage.west);
\draw[flow] (analyzer.east) -- (fingerprint.west);
\draw[flow] ([yshift=-0.32cm]analyzer.east) -- (cost.west);
\end{tikzpicture}}
\caption{\tool{} implementation architecture.}
\label{fig:wgpulens-architecture}
\end{figure*}
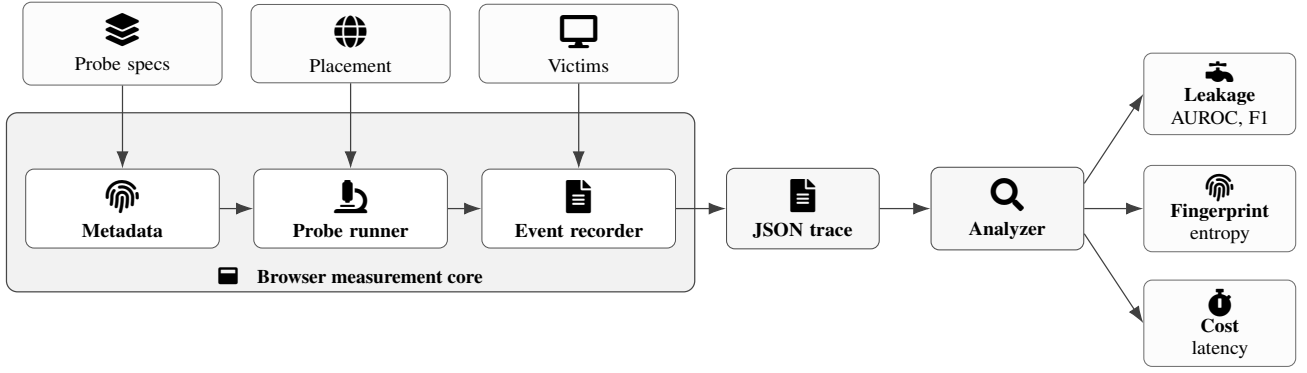

\subsection{Surfaces, Placement, and Observables}

\tool{} measures three surface families. Static exposure covers WebGPU
availability, adapter features, limits, formats, shader capabilities,
timestamp-query availability, WebGL/WebGL2 properties, canvas signatures, and
adapter metadata. Active contention stresses memory, bandwidth, textures,
functional units, atomics, barriers, queue progress, visibility policy, and
scheduling. Persistent state focuses on shader and pipeline compilation, where
cold/warm behavior can reveal previous pipeline-family reuse.
Table~\ref{tab:surfaces-tasks} gives the full surface-to-impact map.

Controlled experiments place the victim at explicit boundaries: same page,
cross-origin iframe, cross-tab, profile/private mode, cross-browser, or
browser/native. Participant runs are same-page measurements for deployment and
fingerprinting; cross-tab leakage requires controlled labels and placement.

The collector supports three observation modes. High-resolution
timing serves as a calibration mode. Coarse browser observables include
\texttt{requestAnimationFrame} boundaries, completed batch counts, queue
completion promises, buffer mapping completion, frame drops, and duration
buckets. Timerless measurements keep only orderings or categorical outcomes.

\subsection{Tasks and Metrics}

The evaluation instantiates four tasks. Pipeline-state inference distinguishes
cold and warm shader or pipeline creation under browser, profile, and cache
boundaries. Native GPU activity inference classifies activity or phase while
the WebGPU probe is co-resident with a controlled native GPU workload.
Controlled workload classification
distinguishes synthetic web workload families or render phases. Participant
fingerprintability measures how WebGPU static exposure and active probes reduce
anonymity sets beyond classic browser, canvas, and WebGL features.
Table~\ref{tab:measurement-tasks} lists the datasets and metrics.

We define endpoints before reporting results. Primary endpoints support the
main claims: pipeline cold/warm boundary AUROC, M1 cross-browser
pipeline AUROC against permutation, browser/native active/idle macro-F1,
participant WebGPU entropy, crawl exposure counts, and pipeline salting
leakage/cost. Secondary endpoints explain heterogeneity; exploratory endpoints
include small pairwise phase contrasts, pilot timer clamps, and dispatch caps.

Binary leakage tasks use AUROC. Multiclass activity and phase tasks use
accuracy and macro-F1. Repeated runs assess stability; bootstrap intervals
quantify uncertainty where available. Label-permutation controls test
closed-world classifiers under randomized labels. Participant fingerprinting
uses uniqueness, collision rate,
entropy, anonymity-set size, and repeat-run linkability. Mitigations report
leakage reduction together with compilation latency, benchmark runtime,
compatibility failures, or frame-level overhead.

\subsection{System Design and Implementation}
\label{sec:wgpulens-design}

\tool{} implements the methodology as a reusable browser/GPU measurement
pipeline. The shared browser core collects metadata, initializes WebGPU, WebGL,
and canvas contexts, executes probe batches, records progress events, and emits
JSON traces. Controlled wrappers add victim placement and labels; participant
wrappers add consent, upload, and repeat-run metadata. Figure~\ref{fig:wgpulens-architecture}
shows the flow from probe specifications to analyzer outputs.

Probes are generated from compact specifications that record surface, access
pattern, workgroup size, dispatch shape, arithmetic mix, memory footprint,
synchronization mode, observable mode, repetition count, and output condition.
Generated WGSL must validate, stay in bounds, avoid uninitialized reads and
intentional data races, and produce deterministic outputs where possible.

Controlled victims are synthetic and labeled. Web victims include static pages,
canvas animation, WebGL render loops, WebGPU render and compute loops,
video-like rendering, map/tiles-style workloads, memory streaming,
texture-heavy rendering, local ML-style kernels, and pipeline compilation.
Native victims include OpenCL on Windows and Metal on Apple platforms. Labels
cover workload family, active/idle state, phase, resource, intensity,
visibility, boundary, and coarse intervals. They are evaluator-only ground
truth; the attacker page observes only its own WebGPU-visible behavior.

The JSON trace is the interface between collection and analysis. Controlled
traces include victim labels and placement metadata; participant traces add
upload metadata and a participant-scoped repeat identifier. The analyzer
reconstructs runs, decodes payloads, joins chunks, validates schema, separates
compatibility failures from completed runs, deduplicates participant-scoped
submissions, and extracts task-specific features. Controlled outputs feed
leakage, boundary, and mitigation tables. Participant outputs feed deployment,
compatibility, entropy, anonymity-set, and repeatability tables.

\section{Controlled Leakage Scenarios}
\label{sec:controlled-scenarios}

\subsection{Controlled Setup}
\label{sec:controlled-setup}

\paragraph*{Claim scope.}

Controlled experiments use fixed browser profiles, fixed origins, fixed window
states, selected adapter configurations, labeled victims, and explicit timer modes.
They support boundary, browser/native, controlled workload classification,
pipeline-state, and mitigation claims. Participant and crawl rows supply
deployment, compatibility, and fingerprinting evidence; leakage claims use
labeled victim placement across a selected boundary.

\paragraph*{Host and browser roles.}

The setup separates adapter, browser, and backend roles. Windows is the
same-host adapter comparison: Chrome and Edge run on the AMD/RDNA-3 default
path and on a forced NVIDIA/Lovelace path. Brave adds Chromium-family
application variation on the adapter configurations present in its traces. Firefox adds
browser-family rows with valid WebGPU traces, but without normalized
vendor/model fields in our data-collection pipeline. M1/Metal provides the Apple backend
with Chrome, Brave, Firefox, and Safari/WebKit coverage. Appendix
Table~\ref{tab:lab-platforms} gives the full host/browser matrix.

\paragraph*{Probe families.}

Pipeline-state experiments vary shader family, pipeline kind, cold/warm/mutated state,
browser/profile boundary, adapter configuration, and observable mode. Browser/native
runs coordinate the browser probe with OpenCL on Windows and Metal on M1. Other
probes cover bandwidth, queues, atomics, textures, and render/compute
signatures. Mitigation experiments repeat selected baselines under
source-level key-separation simulations, timer-clamping variants, or
dispatch-cap variants and record cost.

\paragraph*{Repeat checks.}

We rerun the main controlled experiment sets on later days to check whether the measured
patterns persist across sessions. On Windows, the 2026-06-11 repeat follows the
2026-06-04 baseline and contains 528 valid traces across the AMD and NVIDIA configurations,
covering pipeline state, controlled workload classification, and browser/native
experiments. On M1, the 2026-06-11 repeat covers short stability measurements,
cross-browser pipeline measurements, controlled workload classification, and Metal
browser/native experiments. These repeats provide stability evidence for the
controlled claims; Appendix~\ref{app:lab-stability} lists the repeat artifacts.

\subsection{Pipeline Compilation State Leakage}
\label{sec:surface-results}
\label{sec:pipeline-state}

The strongest controlled surface is WebGPU pipeline compilation state: whether
creating a shader module, compute pipeline, or render pipeline behaves like a
fresh compile or a cache hit in the browser/backend stack. That behavior can
reveal earlier activity in a browser, profile, or cache even after the original
workload has finished.

\paragraph*{Pipeline creation can expose persistent browser/GPU-stack state.}

WebGPU creates shader modules and pipelines before submitting GPU work. The
browser may still compile WGSL, translate it through the backend, and consult
shader or pipeline caches before the call returns. We test whether a pipeline
family is cold or warm.

A contention probe checks for another workload using shared resources now. A
pipeline-state probe checks whether related compilation state already exists,
which can expose earlier activity across origins, profiles, browser variants,
or backends. Private-mode cache behavior is still an open browser-policy
question.

\begin{figure}[t]
\centering
\begin{tikzpicture}[
  font=\scriptsize,
  box/.style={draw=black!65, line width=0.35pt, rounded corners=3pt,
    fill=black!3, align=center, inner sep=4.8pt, minimum height=0.82cm,
    text width=2.68cm},
  state/.style={draw=black!65, line width=0.35pt, rounded corners=3pt,
    align=center, inner sep=4.2pt, minimum height=0.78cm, text width=2.05cm},
  flow/.style={-{Latex[length=1.8mm,width=1.25mm]}, line width=0.42pt,
    draw=black!70}
]
\node[box] (wgsl) at (0,0) {WGSL shader\\family};
\node[box] (api) at (0,-1.24) {\texttt{createShaderModule}\\pipeline creation};
\node[box] (stack) at (0,-2.48) {browser backend\\Dawn, wgpu, WebKit};
\node[state, fill=black!2] (cold) at (-1.42,-3.94) {cold path\\compile/lower\\cache insert};
\node[state, fill=black!9] (warm) at (1.42,-3.94) {warm path\\reuse/cache hit\\shorter path};
\node[box] (obs) at (0,-5.40) {attacker observes\\own completion signal};

\draw[flow] (wgsl.south) -- (api.north);
\draw[flow] (api.south) -- (stack.north);
\draw[flow] ([xshift=-0.34cm]stack.south) -- (cold.north);
\draw[flow] ([xshift=0.34cm]stack.south) -- (warm.north);
\draw[flow] (cold.south) -- ([xshift=-0.34cm]obs.north);
\draw[flow] (warm.south) -- ([xshift=0.34cm]obs.north);
\end{tikzpicture}
\caption{Pipeline-state probe.}
\label{fig:pipeline-state-probe}
\end{figure}
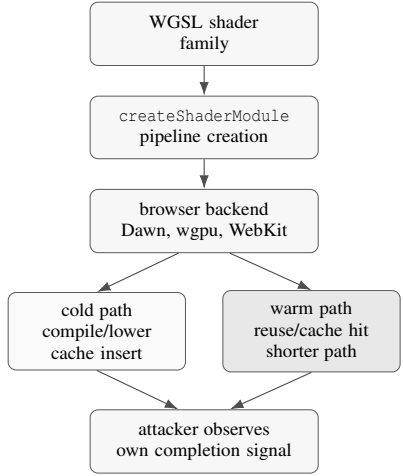

\paragraph*{The probe uses valid WebGPU pipeline creation.}

The pipeline-state experiment uses a controlled cold/warm protocol. We create a
valid WGSL shader or pipeline family under a cold browser, profile, or cache
condition, re-create it under a warm condition, and optionally create mutated or
unrelated families as closely related controls. The attacker records only its own
observable: duration, coarse bucket, completion ordering, or ordinal category.

All shader code is valid WGSL. The experiment uses ordinary pipeline creation
and the attacker's own completion signal: no out-of-bounds access, undefined
shader behavior, driver exploit, or victim memory disclosure.

The controls test simpler explanations such as generic warmup or trial-order
drift. The protocol varies shader family, pipeline kind,
mutated variants, unrelated families, browser profile, origin placement,
adapter configuration, and trial order. Same-origin cross-tab rows are kept as negative
boundary controls, and the M1 cross-browser aggregate is compared against a
label-permutation baseline. These controls constrain the signal without fully
localizing the layer that retains the state, which may live in browser code,
Dawn/wgpu/WebKit policy, driver shader caches, OS filesystem caches, the GPU
process, or a combination of those layers. We therefore attribute the signal
only to the browser/GPU stack, not to a specific browser, backend, driver, or
cache layer.

\paragraph*{Cold/warm state is separable on both Windows adapter configurations.}

The local baseline asks whether a WebGPU page can distinguish a cold pipeline
family from one that has already been compiled. On the Windows hybrid-GPU host,
same-page cold/warm separability is strong under high-resolution JavaScript
timing: AUROC reaches 0.986 on the default AMD/RDNA-3 configuration and 0.961
on the forced NVIDIA/Lovelace configuration. These are calibration rows for the
rest of the pipeline-state study and establish measurability on both adapter
configurations.

The same-host comparison is useful because it holds the CPU, OS, browser
install, and local environment fixed while changing the adapter configuration.
The result is not specific to a single GPU: the browser pipeline-creation path
exposes measurable cold/warm structure on both AMD and forced NVIDIA paths.

The Apple M1 Pro/Metal results show that the surface is not Windows-only. In
the M1 cross-browser experiment set, each browser-pair row uses 8 trace files and
480 high-resolution events split evenly between cold and warm states. Across
the three browser pairs, the aggregate high-resolution AUROC is 0.718 versus a
permuted p95 of 0.522. This is weaker than the Windows same-page rows, but it
provides an independent Apple/Metal backend result.

\paragraph*{Selected placements preserve pipeline-state signals.}

Pipeline-state leakage is boundary-dependent. Table~\ref{tab:pipeline-boundary}
summarizes the primary rows. Same-origin cross-tab rows on Windows are near
chance in the short-interval setting, while selected cross-origin, HTTPS
iframe, cross-profile, Chromium-family application, and M1 cross-browser rows
remain positive.

\begin{table*}[t]
\centering
\caption{Primary pipeline-state boundary rows. Evidence reports the sample
size, repeat, or permutation control available in the derived artifact.}
\label{tab:pipeline-boundary}
\footnotesize
\setlength{\tabcolsep}{3pt}
\renewcommand{\arraystretch}{0.95}
\begin{tabular}{@{}p{0.13\textwidth}p{0.26\textwidth}p{0.31\textwidth}r p{0.12\textwidth}@{}}
\toprule
Role & Configuration & Evidence & AUROC & 95\% CI \\
\midrule
Calibration & Win AMD, same-page, Chrome, high-res & 3 files; 90/90 cold/warm; repeat 0.965 & 0.986 & [0.961--0.999] \\
Calibration & Win NVIDIA, same-page, Chrome, high-res & 3 files; 90/90 cold/warm; repeat 0.933 & 0.961 & [0.929--0.985] \\
\midrule
Web boundary & Win AMD, cross-origin, high-res & 3 files; 90/90 cold/warm; perm. p95 0.581 & 0.948 & [0.907--0.981] \\
Web boundary & Win AMD, HTTPS iframe, high-res & 3 files; 90/90 cold/warm; artifact aggregate & 0.918 & [0.872--0.958] \\
Profile boundary & Win AMD, Chrome cross-profile, high-res & 3 files; 90/90 cold/warm; perm. p95 0.581 & 0.958 & [0.927--0.982] \\
Negative control & Win AMD, same-origin cross-tab, high-res & 3 files; 90/90 cold/warm; repeat 0.507 & 0.545 & [0.491--0.623] \\
Negative control & Win NVIDIA, same-origin cross-tab, high-res & 3 files; 90/90 cold/warm; repeat 0.508 & 0.506 & [0.471--0.596] \\
\midrule
Chromium-family, small-\(n\) & Win Brave $\rightarrow$ Chrome, high-res & 1 file; 20/20 cold/warm & 0.996 & [0.985--1.000] \\
Chromium-family, small-\(n\) & Win Chrome $\rightarrow$ Brave, high-res & 1 file; 20/20 cold/warm & 0.999 & [0.993--1.000] \\
M1 X-browser & M1 Brave $\rightarrow$ Chrome, high-res & 8 files; 240/240 cold/warm; perm. p95 0.542 & 0.744 & [0.700--0.788] \\
M1 X-browser & M1 Firefox $\rightarrow$ Chrome, high-res & 8 files; 240/240 cold/warm; perm. p95 0.543 & 0.712 & [0.666--0.758] \\
M1 X-browser & M1 Chrome $\rightarrow$ Brave, high-res & 8 files; 240/240 cold/warm; perm. p95 0.537 & 0.700 & [0.653--0.747] \\
\bottomrule
\end{tabular}
\vspace{0.3em}
\begin{minipage}{0.96\textwidth}
\footnotesize \emph{Notes.} CIs are 95\% intervals: bootstrap over cold/warm
samples where event vectors are retained, and analytic AUROC intervals for M1
rows whose derived artifacts retain counts and permutation controls but not
event vectors. For \(A \rightarrow B\) rows, \(A\) creates or warms the
pipeline family and \(B\) performs the measured probe.
\end{minipage}
\end{table*}

The boundary determines the claim. Same-page rows are surface discovery;
cross-origin or iframe rows test web isolation; cross-profile rows test
browser-state isolation. The high Windows Brave/Chrome rows are
Chromium-family cache-policy results: the browsers differ at the application
layer but share enough implementation, backend, driver, or cache behavior to
preserve the signal. With 20 observations per direction, they are small-\(n\)
corroboration, not precise effect-size estimates. The M1 rows provide an
Apple/Metal cross-browser check. Only Firefox$\rightarrow$Chrome crosses
rendering-engine families; Brave$\leftrightarrow$Chrome is a cross-application
comparison within the Chromium family.

The table supports controlled placement claims. Origin-only causality requires
another factorial experiment: the current design does not isolate origin as the
sole causal factor. Origin, tab placement, process placement, and cache scope
vary together in these comparisons.

\paragraph*{Controlled repeats preserve the pattern.}

We reran the primary pipeline rows on later collection days on both
controlled platforms. The pattern is stable: Windows same-page
pipeline state remains strong (AMD 0.986 to 0.965; NVIDIA 0.961 to 0.933),
Windows same-origin cross-tab remains near chance (AMD 0.545 to 0.507; NVIDIA
0.506 to 0.508), and the M1 cross-browser aggregate remains positive (0.718 to
0.706). Safari same-page remains weak (0.536 in the repeat). These rows are
part of a broader controlled multi-day repeat that also covers controlled
workload classification and browser/native activity.

\paragraph*{Pipeline state depends strongly on observation mode.}

The strongest measurements use high-resolution JavaScript timing. On M1, the
cross-browser experiment set reaches aggregate AUROC 0.718 under
high-resolution timing, while coarse-frame measurements fall to 0.512 and
timerless ordinal observations are dominated by ties. Same-browser
M1 rows show the same pattern: Chrome and Brave are positive under
high-resolution timing, while coarse and ordinal modes are near chance.

Timer restrictions can reduce this surface under several deployments. Timer
precision is one axis of the measurement model: pipeline state is
high-resolution-sensitive, while later browser/native rows show stronger
coarse-frame signals.

\paragraph*{Controls and negative rows bound the claim.}

The pipeline-state claim relies on controls as much as on positive rows. The M1
cross-browser aggregate is compared to a label-permutation control: AUROC 0.718
versus a permuted p95 of 0.522. Negative or weak rows remain in the main table,
including Windows same-origin cross-tab rows near chance for both AMD and
NVIDIA.

These controls keep the result precise: pipeline compilation state is strongly
measurable locally, remains positive across selected controlled origin/profile,
Chromium-family, and Apple/Metal cross-browser placements, and weakens sharply
under coarse or ordinal observables in several M1 rows.

\paragraph*{Pipeline state points to cache partitioning.}

Pipeline compilation state matters because it is a persistent-state surface.
If pipeline cache reuse crosses controlled boundaries such as origins,
profiles, or cross-browser placements, a page may infer that related shader
activity has occurred before. Private-mode and restart behavior are important
browser-policy questions for future experiments. The implication is
persistent state inference, not only transient contention.

For this reason, Section~\ref{sec:mitigations} treats pipeline cache
partitioning and source-level key-separation simulations as first-class
policies. The browser question is whether stacks can partition or
normalize pipeline state at acceptable cost while preserving WebGPU
functionality.

\subsection{Inferring Native GPU Activity from \texorpdfstring{\mbox{WebGPU}}{WebGPU}}
\label{sec:boundary-results}
\label{sec:browser-native}

The second controlled result crosses the browser/native boundary. The attacker
is a WebGPU page; the victim is a controlled native GPU workload. The result is
activity and phase inference for synthetic native workloads, without victim
memory or content.

\paragraph*{The boundary is browser WebGPU versus native GPU work.}

The Windows native victim uses OpenCL and exercises repeated compute and phase
patterns under both the default AMD/RDNA-3 path and forced NVIDIA/Lovelace
paths. On Apple, the M1 experiment uses a native Metal victim and a Chrome
WebGPU probe. In both cases, victim labels are evaluator-only ground truth; the
attacker observes only its own WebGPU progress, timing, or coarse completion.
The victim is outside the browser process family, so positive rows indicate
contention or scheduling effects visible through the shared OS, driver,
backend, or GPU stack.

\paragraph*{Synthetic native activity is inferable under controlled co-residency.}

The controlled browser/native rows cover activity and phase classification
outcomes (see Table~\ref{tab:case-b}). Throughput slowdowns are supporting
interference evidence, not classification results; Appendix
Table~\ref{tab:browser-native-throughput} reports the exact rows.

\begin{table}[t]
\centering
\caption{Controlled browser/native activity inference. The rightmost column
reports macro-F1. Evidence reports the sample size and the strongest available
control for the headline row. CIs are accuracy CIs when the derived artifact
reports accuracy intervals.}
\label{tab:case-b}
\footnotesize
\setlength{\tabcolsep}{2.5pt}
\renewcommand{\arraystretch}{1.05}
\begin{tabularx}{\linewidth}{@{}
  >{\raggedright\arraybackslash}p{0.13\linewidth}
  >{\raggedright\arraybackslash}p{0.23\linewidth}
  >{\raggedright\arraybackslash}X
  r@{}}
\toprule
Platform & Task & Evidence and control & Macro-F1 \\
\midrule
Win AMD & Exploratory pairwise native phase & Repeated timer aggregates; 48
samples per pair; supporting pairwise rows & 0.875--0.937 \\
Win NVIDIA & Active/idle & High-resolution aggregate; 96 samples & 0.738 \\
Win NVIDIA & Active/idle & Coarse RAF + barrier; 32 samples & 0.782 \\
M1/Metal & Four-phase native classification & Coarse RAF + texture sampling;
32 samples; accuracy CI [0.813, 1.000]; perm. max F1 0.452 & 0.903 \\
M1/Metal & Active/idle & Coarse RAF aggregate; 96 samples; accuracy CI
[0.917, 0.990]; perm. max F1 0.695 & 0.946 \\
\bottomrule
\end{tabularx}
\end{table}

\paragraph*{M1/Metal is strongest for active/idle inference.}
Unlike the pipeline-state measurements, the strongest M1/Metal results use
coarse frame-level signals. The native
Metal repeat set contains 288 valid Chrome/Apple-Metal traces.
Four-phase classification reaches macro-F1 0.903 in the primary
coarse frame-level timing + texture-sampling row with 32 samples and
accuracy CI [0.813, 1.000]. Active/idle reaches macro-F1 0.946 over the broader
coarse frame-level aggregate with 96 samples and accuracy CI [0.917, 0.990].

\paragraph*{Windows rows are more heterogeneous.}
AMD rows are strongest for exploratory pairwise phase contrasts, reaching
0.875--0.937 macro-F1 across repeated timer aggregates. NVIDIA rows are
stronger for active/idle in the focused repeat,
reaching 0.738 macro-F1 over 96 high-resolution traces and 0.782 macro-F1 in a
coarse frame-level + barrier row. This heterogeneity motivates the
multi-platform design: adapter, backend, driver, and scheduler behavior all
affect performance on the same inference task.

\paragraph*{Controlled repeats keep active/idle as the stable result.}

The browser/native stability repeat preserves the main conclusion. On M1/Metal,
active/idle inference remains the
most stable browser/native result: the 2026-06-11 repeat reaches macro-F1 0.931
for coarse frame-level timing, 0.813 for queue-progress observables, and 0.786
for high-resolution JavaScript timing. The same repeat preserves a strong four-phase
coarse frame-level + texture-sampling result at macro-F1 0.914, but with
only 12 samples; broader four-phase aggregates are weaker. The stable M1 claim
is active/idle inference, with four-phase structure as supporting evidence.

The Windows repeat shows the same pattern. Browser/native activity remains
measurable in controlled OpenCL conditions on both AMD and NVIDIA, but
the strongest phase pairs shift across repeats. This supports the
heterogeneity framing and keeps small pairwise rows out of the headline claim.

\paragraph*{Permutation controls and synthetic victims bound the claim.}

Label-permutation controls keep the result grounded. The Windows focused
four-class native-phase control has max macro-F1 0.454 across 20 seeds. The M1
native-phase control has max macro-F1 0.452, and the M1 active/idle control has
max macro-F1 0.695. Pairwise rows are easier to overfit under permutation, so
aggregate and repeated rows carry the main evidence. The M1 active/idle control
is conservative because the compact repeat set preserves session and experiment
structure; we compare against that higher control.

The victims are synthetic and controlled. The browser-policy result is that a
WebGPU page can observe coarse native GPU activity under controlled conditions.
That is the right unit for evaluating browser and GPU-stack isolation policy.

\subsection{Other WebGPU Resource Surfaces}
\label{sec:use-cases}
\label{sec:resource-workload}

\tool{} also measures memory, bandwidth, functional units, atomics, barriers,
queues, textures, and controlled workload classification. These non-pipeline
results provide context for the main pipeline-state finding: they are
measurable but weaker, noisier, and more platform-specific.

\paragraph*{Non-pipeline contention is measurable but weaker.}

Table~\ref{tab:surface-strength} in the appendix summarizes representative
non-pipeline rows alongside the pipeline rows for scale. On the Windows host,
AMD Chrome reaches AUROC 0.660 for a high-resolution bandwidth row and 0.684
for a queue-progress functional-unit row. Forced NVIDIA Chrome reaches 0.797
for a short-interval queue-progress functional-unit row. Timerless non-pipeline
rows peak at 0.575 locally. These above-chance rows provide supporting coverage;
the primary privacy claim rests on pipeline state.

These rows give a clear ordering: pipeline state is the strongest controlled
surface, browser/native activity is the strongest cross-boundary activity
result, and non-pipeline contention provides coverage and negative evidence for
future mitigations.

\paragraph*{Controlled workload classification is measurable in synthetic settings.}

Controlled render-phase classification is possible in synthetic settings, but
this task is substantially narrower than real-site fingerprinting. The M1
Chrome/Brave Apple/Metal dataset
contains 576 validated traces across three render phases, three probe families,
four timer modes, and eight repeats. The primary high-resolution row reaches
accuracy 0.729 and macro-F1 0.731 with 95\% bootstrap interval [0.660, 0.799].
Ten label-permutation controls over the same dataset have maximum primary
accuracy 0.583.

The Windows workload rows show the same caution. Controlled WebGPU
render-phase workloads can be distinguished, with strength depending on adapter
and observable. Realistic WebGL/canvas-style victims are near chance in the
current pilot tests. We use this result as supporting evidence for controlled
workload-structure measurement; open-world website fingerprinting requires a
separate experiment.

\paragraph*{Controlled repeats preserve the moderate workload claim.}

Repeated experiments preserve this moderate interpretation. The hidden label
is a synthetic WebGPU workload or render phase. On M1/Metal, the 2026-06-11
repeat classifies controlled render phases with macro-F1 0.740 under
\texttt{coarse\_raf}, 0.700 under
\texttt{high\_res\_js}, and 0.696 under \texttt{queue\_progress}. The
\texttt{timerless\_ordinal} row remains below baseline at macro-F1 0.213. On
Windows, the repeated workload-classification experiment remains strong on the AMD path
(macro-F1 0.963 for both \texttt{high\_res\_js} and \texttt{coarse\_raf}) and
weak on the forced NVIDIA path. This repeat supports a controlled workload
classification claim.

\paragraph*{Timer effects are surface-specific.}

The non-pipeline rows reinforce the timer results from
Section~\ref{sec:pipeline-state}. Timer restrictions help, but their effect is
surface-specific: some scheduler rows collapse under coarse timing, some coarse
browser/native rows remain strong, and timerless ordinal rows are weak in the
current Windows non-pipeline dataset. Timer policy is one dimension of a broader
measurement problem.

\section{Participant Field Measurements}
\label{sec:field-fingerprintability}

The opt-in participant field study measures three things: deployment and
compatibility, active WebGPU diversity and sample-level fingerprintability, and
repeatability through repeated local-storage identifiers.

\subsection{Participant Coverage and Compatibility}

\paragraph*{The participant study is opt-in and same-page.}
The participant field study uses an opt-in benchmark deployed as a public page.
The page presents a short disclaimer, asks for consent, runs after one button
press, shows progress, and uploads a compressed JSON payload through a form
endpoint. The participant protocol is same-page by design, with no victims, extra tabs, or
native workload coordination. It records classic browser properties, canvas
hashes, WebGL/WebGL2 capabilities, shader precision, extension counts, render
hashes, WebGPU adapter features and limits, active WebGPU outputs, compact
active timing summaries, dynamic same-page probe summaries, completion status,
and a random localStorage-scoped identifier for soft deduplication. That
identifier is not a login, IP address, account identity, browsing history, or
local-file identifier.

\paragraph*{Participant uploads reconstruct into completed records.}
Large gzip/base64 payloads are chunked across participant-upload rows, so we
reconstruct logical browser runs before analysis. The export contains 7,999
participant-upload rows, 1,534 reconstructed logical runs, 1,127 completed
runs, 1,095 deduplicated completed records, and 80 local-storage identifiers
seen more than once; Appendix~\ref{app:participant-ingestion} gives the funnel.
The 118 decode errors are incomplete chunk groups, not malformed completed
payloads, and are excluded from primary metrics.

\paragraph*{Compatibility follows browser and adapter exposure.}
Compatibility failures are part of the result: a browser where
\texttt{navigator.gpu} is unavailable or \texttt{requestAdapter} returns null
is counted as an unavailable WebGPU configuration. Completion varies by browser: Chrome
completes 678 of 898 decoded runs, Safari
151/211, Edge 115/132, Firefox 84/137, Chromium-family variants 64/112, and
other browsers 35/44. The strongest split is by adapter exposure. When a
concrete GPU vendor is exposed, completion is high: Intel 402/413, AMD 137/139,
NVIDIA 123/124, Apple 297/297, Qualcomm 42/57, and ARM 35/35. Unknown adapters
complete only 84 of 462 decoded runs.
WebGPU exposure depends on browser, platform, and adapter configuration. The
second campaign expands mobile and Apple coverage: the completed set includes
145 iOS runs and 78 Android runs.

\subsection{Active WebGPU Signals and Fingerprintability}

\paragraph*{Signatures separate static exposure, active outputs, and timing.}
We compute signatures for four families of features. Classic signatures include
browser, OS, version, screen, timezone, hardware, fonts, canvas, and WebGL.
WebGPU static signatures use support, adapter features, limits, and metadata
that the browser exposes before active work. WebGPU active signatures are
reported in two layers: output-only render/compute/pipeline hashes, and an
active WebGPU signature that adds stability flags and compact duration
summaries.
Dynamic same-page summaries capture coarse behavior from the short WebGPU probe
suite.

\paragraph*{Analysis signatures exclude upload and recruitment identifiers.}
Before computing anonymity sets, the analyzer canonicalizes feature
objects and excludes fields that would trivially identify a run: participant
identifier, local run index, upload filename, chunk count, form row metadata,
collection or completion timestamps, completion code, participant-platform
completion metadata, and recruitment URL parameters. The standalone active
WebGPU signature is built from deterministic render, compute, and pipeline
outputs, within-run stability flags, and compact duration summaries for
pipeline creation, submission, and readback. The dynamic signature is reported
separately and comes from the short same-page probe suite.

\paragraph*{The participant protocol runs the same active probes without labels.}
The participant benchmark reuses the browser-side probes used by controlled
scenarios: render work, compute work, pipeline creation, compact timing
summaries, and short dynamic same-page probes. In the participant protocol these
measurements are unlabeled field observations. They support compatibility,
active-signal diversity, fingerprintability, and repeatability across real
stacks; controlled scenarios add labels and placement control.

\paragraph*{Active WebGPU is distinctive within this sample.}
The full classic fingerprint is nearly saturated in this export: 1,091 of 1,095
completed deduplicated records are unique. The main field result is standalone
WebGPU fingerprintability. Static exposure is coarse; output hashes distinguish
some graphics classes; the active WebGPU signature combines output hashes,
stability flags, and compact timing summaries. Table~\ref{tab:webgpu-static-active}
shows the ablation. Distinctiveness and repeat stability are different
properties: timing-rich signatures are highly distinctive in this sample, while
repeated local-storage identifiers show more limited exact stability.

\begin{table}[t!]
\centering
\caption{Standalone WebGPU anonymity sets over 1,095 completed deduplicated
participant records. Output-only rows use render, compute, and pipeline hashes
without active timing summaries.}
\label{tab:webgpu-static-active}
\footnotesize
\setlength{\tabcolsep}{4pt}
\begin{tabular}{@{}p{0.42\linewidth}rrr@{}}
\toprule
WebGPU signature & Sets & Entropy bits & Largest set \\
\midrule
Static exposure & 122 & 5.188 & 168 \\
Render output hash & 9 & 2.548 & 329 \\
Compute output hash & 14 & 3.219 & 238 \\
Pipeline output hashes & 1 & 0.000 & 1,095 \\
Render + compute + pipeline outputs & 16 & 3.250 & 238 \\
Active signature (outputs + timing) & 1,095 & 10.097 & 1 \\
\bottomrule
\end{tabular}
\end{table}

Static WebGPU fields and output hashes still leave collisions: render and
compute hashes separate broad classes, while the pipeline output hash is stable
across all completed runs. The active WebGPU signature places every completed
deduplicated record in this sample in a singleton anonymity set, including
major subsets: Windows 650/650, macOS 193/193, iOS 133/133, Android 77/77,
Chrome 669/669, Safari 137/137, and known-GPU rows 1,011/1,011. This comes
from WebGPU behavior itself, without the full classic browser fingerprint.

\begin{table}[t!]
\centering
\caption{Sequential participant fingerprinting ablation over completed
deduplicated participant records.}
\label{tab:field-entropy-ablation}
\footnotesize
\setlength{\tabcolsep}{3pt}
\begin{tabular}{@{}lrrrr@{}}
\toprule
Features & Sets & Entropy & \(\Delta\) bits & Singleton fraction \\
\midrule
Browser + OS & 16 & 2.961 & -- & 0.000 \\
+ browser/OS versions & 80 & 4.280 & 1.319 & 0.029 \\
+ screen/timezone/hardware & 901 & 9.658 & 5.378 & 0.721 \\
+ canvas & 938 & 9.750 & 0.092 & 0.772 \\
+ fonts & 977 & 9.833 & 0.083 & 0.831 \\
+ WebGL metadata & 1027 & 9.950 & 0.117 & 0.899 \\
+ WebGL render hashes & 1028 & 9.952 & 0.002 & 0.900 \\
+ WebGPU static & 1031 & 9.961 & 0.009 & 0.903 \\
+ WebGPU active & 1,095 & 10.097 & 0.136 & 1.000 \\
\bottomrule
\end{tabular}
\end{table}

\paragraph*{Active behavior adds entropy over coarse baselines.}
The incremental value of WebGPU depends on the baseline. Against the full
classic fingerprint, the increment is very small because the baseline is already
nearly saturated at 1,091 anonymity sets and 10.089 bits, close to
\(\log_2(1{,}095)=10.097\). This is common in rich browser-fingerprinting
datasets. Over coarser baselines, the active WebGPU signature adds 7.136 bits
to browser/OS buckets, 5.817 bits to browser/OS/version buckets, and 4.909 bits
to WebGPU static exposure alone. The last number shows that active execution
behavior carries information beyond adapter features and limits.

Table~\ref{tab:field-entropy-ablation} shows the cumulative feature ablation.
Final increments are small because earlier classic browser fields already make
almost every record unique in this sample. The full classic fingerprint includes
additional classic fields not shown in the sequential ablation;
Table~\ref{tab:field-entropy-ablation} reports the ordered subset used for
feature-attribution analysis. Entropy and anonymity-set values are empirical
plug-in estimates over the 1,095 completed deduplicated records; we use them to
compare feature sets within this export. Broader uniqueness and tracking claims
would require a planned longitudinal sample.

\subsection{Participant Stability and Repeats}

\paragraph*{Local-storage identifiers provide a repeat-visit check.}
We use a first-party local-storage identifier to control repeated submissions.
The study page stores a random identifier in local storage and includes it in
later uploads from the same browser storage state. In the export, 125 completed
runs carry a local-storage identifier already seen before. Those repeats map to
80 identifiers because some identifiers appear in three or more completed runs.

\paragraph*{Deterministic components are stable within a run.}
All 1,127 completed runs report within-run stable deterministic WebGPU
components: render hash present in 1,127/1,127 runs, compute hash present in
1,127/1,127, pipeline hash present in 1,127/1,127, and the active stability flag
true in 1,127/1,127. This supports an intra-run claim for the deterministic
output components. The active signature, including compact timing summaries,
is used for empirical fingerprintability over completed deduplicated records
(see Table~\ref{tab:field-repeatability}).

\paragraph*{Repeated local-storage identifiers give a limited stability check.}
For rows with denominator 80, stable means that every completed run from the
same first-party local-storage identifier produced the same signature. WebGPU static
signatures are stable in 80/80 repeated identifiers, WebGPU active signatures in
57/80, classic signatures in 62/80, and combined signatures in 45/80. The
1,127/1,127 row supports within-run stability. Rows with denominator 80 show
repeat stability: how often the same local-storage state produced the same
signature in later completed runs. Planned follow-up across time, browser
restarts, browser updates, and driver or OS changes is needed for longitudinal
tracking.

\begin{table}[t]
\centering
\caption{Repeatability checks in the participant dataset. Rows with denominator
80 count local-storage identifiers seen more than once, not people or
individual runs.}
\label{tab:field-repeatability}
\footnotesize
\setlength{\tabcolsep}{5pt}
\begin{tabular}{@{}lr@{}}
\toprule
Check & Stable / comparable \\
\midrule
Intra-run deterministic active components & 1,127/1,127 \\
Repeated WebGPU static signature & 80/80 \\
Repeated WebGPU active signature & 57/80 \\
Repeated classic signature & 62/80 \\
Repeated combined signature & 45/80 \\
\bottomrule
\end{tabular}
\end{table}

\paragraph*{Participant results support deployment, active diversity, and repeatability.}

The participant results show that the harness runs on real browsers,
compatibility failures are structured, active WebGPU probes are high-entropy,
and active components are stable within a run.

\section{Page-Load WebGPU Exposure}
\label{sec:webgpu-exposed-surface}

The crawl asks whether public pages touch WebGPU during normal page load, and
in what source context. We answer it with a Tranco top-10k crawl.
The claim is page-load exposure of adapter probing and WebGPU-bearing source
code; leakage and fingerprintability come from controlled rows and participant
runs.

\paragraph*{The crawl measures page-load exposure.}

The crawler loads each site in Chromium and writes one bounded JSONL record per
URL. It instruments selected WebGPU API calls, records bounded call stacks, scans fetched
scripts for WebGPU, WebGL, and canvas markers, and stores source URLs and
script hashes. Records omit HTML, JavaScript bodies, screenshots, browser
profiles, account data, and browsing histories. The crawl is page-load only:
no clicks, logins, attempts to complete consent dialogs, or application paths,
so it is a lower
bound on interactive WebGPU use. Appendix~\ref{app:crawl-details} defines the
static-reference, adapter-probe, device-probe, and observed-work evidence
levels.

\begin{table}[t!]
\centering
\caption{Page-load WebGPU exposure in the Tranco top-10k crawl. Counts are
conservative; static and dynamic evidence can overlap.}
\label{tab:tranco-crawl-summary}
\footnotesize
\setlength{\tabcolsep}{3pt}
\begin{tabular}{@{}p{0.43\columnwidth}r p{0.34\columnwidth}@{}}
\toprule
Outcome & Count & Interpretation \\
\midrule
Attempted URLs & 10,000 & Crawl scope. \\
Successful page-load records & 7,477 & Pages that reached the measurement point. \\
WebGPU-positive records & 56 & Static or dynamic WebGPU evidence. \\
Dynamic WebGPU-call evidence & 33 & Runtime calls observed during page load. \\
Static WebGPU evidence & 29 & WebGPU tokens in fetched scripts. \\
Adapter-probe-only records & 32 & Environment or capability probing. \\
Device-probe records without work & 1 & Device request observed, but no submitted
work. \\
Static-reference-only records & 23 & Bundled or dormant WebGPU support code. \\
Observed page-load WebGPU work & 0 & No shader, pipeline, queue, query, or map
activity observed. \\
\bottomrule
\end{tabular}
\end{table}

\paragraph*{Page-load exposure is mostly adapter probing and static support code.}

The crawl attempted 10,000 URLs, obtained 7,477 successful page-load records, and
found 56 WebGPU-positive records: 33 with dynamic WebGPU calls and 29 with
static WebGPU evidence. Static and dynamic evidence can overlap. The positives
consist primarily of adapter probes and static references. We observed no page-load compute
workloads: no shader-module creation, pipeline creation, queue submission,
query-set use, or buffer mapping. The most common runtime behavior is
\texttt{navigator.gpu.requestAdapter}; one record reached
\texttt{requestDevice} without submitting work (see
Table~\ref{tab:tranco-crawl-summary}).

\paragraph*{Source context contains fingerprinting-adjacent and capability markers.}

We classify source context conservatively because one crawled bundle can mix
fingerprinting, anti-abuse, analytics, graphics, and capability code. The
largest group contains fingerprinting-adjacent source markers: 28 records
contain WebGPU references near fingerprinting, WebGL, canvas readback, buffer
readback, shader, or identifiers and strings containing fingerprinting-related
names such as \texttt{fp}. Six records are anti-abuse or trust-safety context, six are
media or player capability checks, four are analytics or generic capability
checks, three combine media/player code with neighboring risk markers, and nine
remain manual-review or unclear cases.
Appendix~\ref{app:crawl-details} lists the triage rules and representative
source families.

\paragraph*{Representative positives reproduce across Chromium-family browsers.}

A focused reproduction pass on 13 representative positives, repeated three
times across Chromium-family targets, reproduced runtime WebGPU calls on 12
targets; the remaining target stayed negative. Across Chrome, Edge, and Brave,
outcomes and call counts were stable across the three repetitions, and 11
targets retained the same callsite family. This supports a narrow deployment
claim: adapter probing appears in crawled scripts and is reproduced across the
tested Chromium-family browsers.

\paragraph*{The crawl gives browsers concrete exposure points.}

The crawl connects the controlled and participant evidence to real web pages.
Some crawled scripts query WebGPU as an environment signal, mostly through
adapter probing and bundled support code, with no page-load shader, pipeline,
queue, query, or map activity. This gives browser vendors a concrete place to
evaluate metadata exposure, adapter availability, permission, visibility
rules, and source partitioning.

\section{Mitigation: Pipeline Cache-Key Separation}
\label{sec:mitigations}

Controlled results point to pipeline cache policy: the signal comes from
cold/warm reuse, calling for cache partitioning or source-level key separation.
The key mitigation
question is what browser-controlled state should be partitioned, bucketed,
delayed, perturbed, or gated, and at what cost.

\paragraph*{Mitigations target different WebGPU surfaces.}
The prototype covers five defense families: source partitioning for pipeline
cold/warm state, metadata bucketing for adapter exposure, observable padding
for queue and completion timing, readback perturbation for active output
fingerprints, and WebGPU gating for the strictest mode. The main measured
policy is source-level key separation as a proxy for pipeline cache
partitioning; Appendix~\ref{app:mitigation-supplement} lists the full defense
taxonomy.

\paragraph*{The prototype demonstrates policy choices, not browser patches.}
The extension prototype lets the same stress page run with defenses enabled or
disabled, records active defenses, and shows what each policy changes. Measured
source-level key-separation rows are policy simulations and extension-level
demonstrations; browser-engine cache-key patches are the production target. In
these experiments, modifying the WGSL source prevents reuse of an identical
source-derived key across the tested boundary, approximating cache-key
separation. This is closer to cache busting than a production browser
implementation. A production implementation would attach
origin, top-level site, profile, private-mode, or session state to an internal
browser/backend cache key and measure the same leakage/cost tradeoff.

\paragraph*{Measured rows compare leakage reduction with cost.}
Table~\ref{tab:mitigation-pilots} reports rows with both a baseline and a
mitigated condition. We evaluate each defense as leakage reduction plus cost:
compile latency, frame latency, jank, compatibility loss, correctness loss, or
loss of WebGPU functionality. Source-level key separation is the measured case study
for pipeline cache policy; timer and dispatch rows are exploratory pilots.

\begin{table}[t]
\centering
\caption{Measured pipeline-cache mitigation rows. Source-level key separation
approximates pipeline-cache partitioning as the main case study. Evidence reports the sample
size or pilot status carried by the source artifact. Mitigation
baselines are matched to their mitigation harnesses and may differ from the
primary boundary rows in Table~\ref{tab:pipeline-boundary}.}
\label{tab:mitigation-pilots}
\scriptsize
\setlength{\tabcolsep}{2pt}
\renewcommand{\arraystretch}{0.95}
\begin{tabular}{@{}p{0.14\linewidth}p{0.24\linewidth}p{0.23\linewidth}p{0.20\linewidth}p{0.12\linewidth}@{}}
\toprule
Surface & Policy or condition & Evidence & AUROC base$\rightarrow$mit. (delta) & Cost \\
\midrule
Pipeline & Windows source-key separation & 2 traces; small-\(n\) policy row & 0.994$\rightarrow$0.606 (0.388) & +6.05 ms warm compile \\
Pipeline & M1 source-key separation & 10 files; 150/150 cold/warm per condition & 0.759$\rightarrow$0.520 (0.239) & +0.53 ms warm compile \\
Pipeline & Chromium source partition & 30-trial cross-origin prototype & 0.648$\rightarrow$0.525 (0.123) & extension-level shim \\
\bottomrule
\end{tabular}
\end{table}

\paragraph*{Pipeline cache policy is the clearest measured direction.}
On Windows, the small-\(n\) policy-simulation row reduces cold/warm
AUROC by 0.388 at +6.05 ms warm-compile cost. On M1, the larger M1 experiment
reduces AUROC by 0.239, moves the mitigated result below the
95th-percentile permutation threshold, and adds +0.53 ms warm-compile overhead.
The extension prototype supports the same direction at page level: AUROC drops
from 0.648 to 0.525 in a local cross-origin priming row. These rows support a
direct browser question: should shader and pipeline caches be reused across
origin, profile, session, or private-browsing boundaries?

\paragraph*{Timer and dispatch policies are exploratory.}
Timer clamping is mixed: a 32 ms clamp collapses one scheduler row and weakens
one bandwidth row, while an 8 ms clamp gives inconsistent separability.
The current dispatch-cap experiments validate the instrumentation but do not
show a conclusive mitigation effect. Timer clamps and dispatch caps remain
useful only when tied to a specific surface and cost.
Appendix~\ref{app:mitigation-supplement} lists the exploratory rows.

\paragraph*{Production defenses belong where browsers control the surface.}
Pipeline state needs shader and pipeline cache policy: origin/profile
partitioning, private-mode clearing, or internal cache-key separation. Static
exposure can be reduced by limiting or bucketing the features and limits
exposed to pages. Timing surfaces need timestamp-query
and completion semantics measured against latency and jank. Active output
fingerprints need readback budgets or explicit privacy modes because changing
readback data affects correctness.

\paragraph*{Mitigations need surface-specific testing.}
A WebGPU privacy test suite should implement a
candidate policy, rerun the same rows, and report leakage reduction with
user-visible cost. For the current data, pipeline cache partitioning is the
strongest measured direction. Timer clamping, metadata bucketing, readback
perturbation, and access gating remain useful when their surface and cost are
explicit.

\section{Discussion and Limitations}
\label{sec:discussion-limitations}

\paragraph*{Browser policy.}
WebGPU privacy review should carry the WebGL lesson into compute. WebGL exposed
adapter identity, rendering output, and graphics-stack variation. WebGPU adds
programmable workloads, pipeline creation, queue progress, and backend
scheduling. In our controlled matrix, persistent pipeline state is the clearest
place to start: shader and pipeline cache reuse should be tested across origin,
top-level site, profile, session, private-mode, restart, and application
boundaries. Any policy must report both leakage reduction and user-visible cost,
especially compile latency and compatibility for applications that rely on
pipeline reuse~\cite{webgpu-dispatch-overhead,webllm,llamaweb}.

\paragraph*{What browsers should test.}
The same policy question should be applied to each WebGPU state separately:
adapter metadata, shader-module state, pipeline-cache state, queue completion,
buffer mapping, readback output, and background scheduling. For each state,
browsers should vary the boundary, the observable, and the cost. The relevant
boundaries include same page, cross-origin placement, tab, profile, private
session, browser application, restart, and native co-residency. The relevant
observables include high-resolution timing, frame-level events, queue or map
completion, output hashes, and ordinal outcomes. The relevant costs include
compile latency, frame pacing, jank, fallback behavior, and correctness for
applications that expect exact GPU readback. This is why timer clamping,
metadata bucketing, cache partitioning, readback perturbation, and access gating
cannot be evaluated as one defense.

\paragraph*{Evidence boundaries.}
Table~\ref{tab:evidence-design} defines what each dataset can support.
Controlled rows support leakage, boundary, browser/native, workload, and
mitigation claims because they include labels and explicit placement.
Participant rows support compatibility, heterogeneity, sample-level
fingerprintability, and repeated local-storage identifiers on deployed browsers.
Page-load crawl rows support WebGPU exposure and source context on public pages.
These roles keep controlled leakage, field distinctiveness, repeat behavior,
and page-load exposure from being treated as the same claim.

\paragraph*{External validity.}
The controlled victims are synthetic by design: they provide clean labels and
avoid real user activity, but they do not cover all real workloads. Workload
classification covers controlled render-phase inference; browser/native rows
cover controlled native activity and phase inference. Windows AMD/NVIDIA and
Apple M1/Metal are the strongest controlled experiment sets. Mobile, Safari, Intel,
Qualcomm, and ARM coverage remain collection targets. Participant repeats show
the same first-party local-storage state returning to the benchmark; planned
follow-up is needed for stability across browser restarts, software updates,
driver changes, and longer time windows
~\cite{hiding-crowd,browser-fp-survey,fp-scanner,fingerprint-dynamics,
long-term-fp,fp-radar}.

\paragraph*{Root cause and instrumentation.}
Browser GPU processes, Dawn/wgpu/WebKit policy, OS scheduling, drivers,
filesystem caches, thermal state, and power policy can all affect the measured
signals. \tool{} records metadata and supports repeats, but full attribution
requires browser and driver instrumentation. A platform test suite can make that
work tractable: hold the workload, boundary, observable, and metric fixed while
browsers, backends, drivers, and mitigation policies change
~\cite{w3c-fingerprinting-guidance,w3c-threat-model-web,
w3c-design-principles}.

\section{Related Work}
\label{sec:related-work}

\textbf{Web graphics fingerprinting.}
WebGPU extends a long web-graphics privacy line. Panopticlick introduced
entropy and anonymity-set metrics~\cite{panopticlick}. Canvas/WebGL, OS,
hardware, and graphics features distinguish browsers
~\cite{pixel-perfect,cross-browser-fp}, and DrawnApart showed that WebGL
execution behavior can identify GPUs~\cite{drawnapart}. WebGL also shaped
mitigation and policy: Rendered Private rewrites GLSL, while
\texttt{WEBGL\_debug\_renderer\_info} is treated as privacy-sensitive adapter
exposure~\cite{rendered-private,khronos-webgl-debug-renderer-info,
mdn-webgl-debug-renderer-info}. \tool{} measures the newer WebGPU surfaces:
programmable workloads, queue progress, and pipeline state.

\textbf{WebGPU attacks and fingerprinting.}
WebGPU-SPY applies cache occupancy on Intel integrated GPUs to website
fingerprinting~\cite{webgpu-spy}. Giner et al. automate drive-by WebGPU cache
attacks across desktop GPUs and vendors~\cite{giner-webgpu-cache}.
AtomicIncrement and LockedApart use WebGPU or compute scheduling for device
fingerprinting~\cite{atomicincrement-webgpu,lockedapart}. \tool{} complements
these attacks with a measurement design that compares persistent compilation
state, browser/native co-residency, non-pipeline probes, timer-degraded
observations, participant field measurements, page-load adapter probing, and
mitigation cost.

\textbf{Graphics and GPU side channels.}
GPU side channels predate WebGPU. Prior work covers page-content exposure from
rendering vulnerabilities~\cite{stealing-webpages-gpu}, web-exposed graphics
and compute channels~\cite{grand-pwning-unit,rendered-insecure},
context-switching leakage~\cite{leaky-dnn}, integrated CPU-GPU channels
~\cite{leaky-buddies}, and multi-GPU contention~\cite{spy-gpu-box}. Browser
graphics also exposes compression~\cite{gpuzip},
frequency/power/temperature effects~\cite{hot-pixels}, SVG filter leakage
~\cite{pixel-thief}, mobile pixel stealing~\cite{pixnapping}, and Apple
M-series cache effects~\cite{exam-mseries}. Timerless GPU caches, uncore
resources, and partitioned deployments add further resource families
~\cite{invalidate-compare,veiled-pathways,behind-bars}. These results motivate
\tool{}'s memory, texture, functional-unit, atomic, queue, scheduler, and
persistent-state probes.

\textbf{Browser timing and shared resources.}
Browser side-channel work shows that ordinary APIs can expose shared-resource
behavior. Timer mediation remains subtle
~\cite{clock-ticking,trusted-browsers,fantastic-timers}; generic computation
can identify devices~\cite{clock-around}; and JavaScript cache attacks remain
practical under restrictive APIs~\cite{spy-sandbox,robust-cache-occupancy,
primeprobe-js0}. OPFS contention~\cite{frost} and extension-policy leakage
~\cite{extension-breakdown} show similar patterns in other browser resources.
Cookies from the Past combines timing, web-scale scanning, validation,
stability tests, and countermeasures~\cite{cookies-past}. \tool{} applies this
measurement style to WebGPU with controlled labels, field traces, crawl data,
and leakage/cost rows.

\textbf{Browser fingerprinting and measurement observatories.}
Large-scale fingerprinting studies provide the field-methodology baseline.
FPDetective, Cookieless Monster, and OpenWPM measured web fingerprinting and
standardized browser automation~\cite{fpdetective,cookieless-monster,openwpm}.
Hiding in the Crowd, surveys, FP-STALKER, FP-Scanner, FP-Inspector, FP-Radar,
fingerprint dynamics, and long-term studies refine uniqueness, stability,
countermeasure consistency, and linkability metrics
~\cite{hiding-crowd,browser-fp-survey,fp-stalker,fp-scanner,fp-inspector,
fp-radar,fingerprint-dynamics,long-term-fp}. Static/dynamic measurement and
emerging-technique work motivate evaluating new browser APIs during deployment
~\cite{web-fp-wild,emerging-fp}. \tool{} measures WebGPU's incremental
fingerprintability beyond these established features. It also brings WebGPU
measurement into the same deployment and field-study frame used by these
observatory efforts.

\textbf{Correctness, memory safety, and mitigation context.}
Recent WebGPU and GPU-stack work studies bugs or correctness failures:
LeftoverLocals demonstrates GPU local-memory disclosure~\cite{leftoverlocals},
Whispering Pixels studies register residual leakage~\cite{whispering-pixels},
and SafeRace, WebGlitch, and DarthShader study WGSL races, randomized WebGPU
programs, and shader translation/compiler pipelines
~\cite{saferace,webglitch,darthshader}. Mitigations such as GPUGuard propose
spatial and temporal partitioning for contention channels~\cite{gpuguard};
W3C guidance, privacy budgets, Brave, Firefox RFP, and Tor Browser illustrate
browser-side capability and fingerprinting policy
~\cite{w3c-fingerprinting-guidance,w3c-design-principles,
w3c-threat-model-web,privacy-budget,brave-fp-protections,firefox-rfp,
tor-fp-protections}. These works provide context for evaluating WebGPU
leakage and browser defenses. \tool{} evaluates WebGPU mitigation as a
surface-specific leakage/cost tradeoff.

\section{Conclusion}
\label{sec:conclusion}

WebGPU brings GPU computation into the web platform. Its validation rules
protect memory safety, but they do not remove every privacy-relevant signal.
The key question is: which state crosses which boundary, under which
observable, and at what cost to mitigate it.

\tool{} answers that question with controlled scenarios, participant field
measurements, and page-load crawl data. Persistent pipeline compilation state is
the strongest measured surface. Browser/native active-idle inference is the
strongest co-residency result under synthetic native workloads. Controlled
workload classification and other resource surfaces are measurable but more
platform-specific. Timer reduction helps some rows, especially pipeline-state
rows, but it does not solve the problem by itself.

Field participant runs show active WebGPU behavior is highly distinctive when
compact timing summaries are included. Deterministic active components are
stable within runs, while exact repeat stability across repeated local-storage
identifiers is lower. The page-load crawl shows WebGPU in public-page scripts
mainly as adapter probing and static support code, not as observed page-load
shader, pipeline, queue, query, or map work.

The first mitigation target is browser-controlled persistent state. Browsers
should test shader and pipeline cache partitioning across origin, top-level
site, profile, session, private-mode, and application boundaries. They should
also measure adapter metadata, completion observables, active output hashes,
and browser/native scheduling separately. Source-level key separation is a
proxy, but it shows the leakage/cost tradeoff that browser cache policy must
make explicit.

\section*{Ethical Considerations}
\label{app:ethics}

Participant measurements are opt-in, synthetic, and local to the study page.
The page explains the research purpose, requests consent, shows progress, and
uploads only the benchmark payload. The payload excludes browsing history, page
contents, account identifiers, keystrokes, contacts, local files, and real
website activity. The collection was not reviewed by an IRB; participant-based
claims are therefore limited to technical compatibility, sample-level
fingerprintability, and repeat behavior observed within the benchmark data.

The participant benchmark uses a random first-party localStorage identifier for
deduplication and repeat checks; it is not a login, IP address, or real
identity. In the CloudResearch deployment, a completion code is shown after
upload and stored in the JSON so completion can be reconciled without platform
participant IDs. Payment, approval, marketplace account state, eligibility,
compensation, and account fields stay outside the released traces and are not
analysis variables. Device targeting is used only to obtain browser and
platform coverage.

The upload mechanism uses Google Forms as transport for compressed JSON
payloads. The benchmark payload excludes IP addresses and account identifiers,
but Google may retain operational logs outside the research artifact. Raw
response exports and decoded high-entropy traces are restricted to the research
team and are not included in the release artifact. Public trace releases bucket
or normalize high-entropy metadata, remove completion and provider metadata,
and separate raw internal review material from reproducible artifacts. Clearing
site storage prevents later runs from being linked through the previous local
identifier; it does not remove submitted records.

The page-load crawl uses public pages only. It does not log in, click through
sites, complete consent dialogs, or interact with application workflows. Crawl
records are bounded page-load WebGPU summaries: call metadata, bounded call
stacks, source URLs, script hashes, and static markers. The artifact excludes
full HTML, full JavaScript bodies, browser profiles, account data, and crawl
screenshots. The runner uses per-page timeouts, a per-URL budget, one
concurrent page load by default, and blocked image, media, and font resources.

\section*{Open Science}
\label{app:artifact}

The anonymous review artifact is available at
\url{https://anonymous.4open.science/r/WGPULens/}. It packages the
\tool{} measurement code, WGSL probes, probe generator, browser runners,
synthetic victims, participant benchmark tooling, schemas, participant
decoders, analysis code, normalized derived data, Chromium and Firefox
extension prototypes, and reproduction instructions. The extension artifact
exposes the same defense families discussed in Section~\ref{sec:mitigations}:
source partitioning, static masking, observable padding, readback noise, and
access gating. The artifact labels these prototype mechanisms ``static
masking'' and ``readback noise''; in the paper, we refer to the broader policy
families as metadata bucketing and readback perturbation. These extension
defenses are released as reproducible
demonstrations and policy experiments; browser-engine isolation remains the
production target. Code that could be repurposed as a turnkey attack against
real websites is restricted or released in a non-operational form.

The artifact is organized around the evidence boundaries used in the paper.
For controlled scenarios, it includes normalized pipeline-boundary, stability,
browser/native, workload, and mitigation summaries sufficient to regenerate the
reported AUROC, macro-F1, confidence-interval, count, and cost rows. For the
participant field study, it includes privacy-reviewed aggregate and
deep-analysis outputs for the ingestion funnel, entropy, anonymity sets, active
WebGPU ablations, and repeat evidence. For the Tranco crawl, it includes
bounded crawl summaries, provenance metadata, and the focused case analysis
used to support the page-load exposure claims. Raw participant exports,
completion metadata, high-entropy identifiers, full crawled JavaScript
bodies, browser profiles, and private lab caches are not included in the
anonymous artifact.

Tiered checks can be run from the artifact root. The health tier verifies
manifest hashes, sample traces, and dependencies. The derived tier regenerates
the packaged summaries for pipeline state, browser/native activity, participant
fingerprintability, page-load WebGPU exposure, and mitigation. Optional tiers
cover local interactive demos and small machine-dependent reruns. These local
reruns are provided for inspection and usability; the paper metrics are
reproduced from the normalized derived data.

\bibliographystyle{plain}
\bibliography{refs}

\appendix
\section{Technical Appendix}
\label{app:methodology-supplement}

\setlength{\textfloatsep}{6pt plus 1pt minus 2pt}
\setlength{\floatsep}{5pt plus 1pt minus 2pt}
\setlength{\intextsep}{5pt plus 1pt minus 2pt}
\setlength{\abovecaptionskip}{3pt}
\setlength{\belowcaptionskip}{1pt}

\begin{table}[H]
\centering
\caption{Measurement tasks, datasets, and metrics.}
\label{tab:measurement-tasks}
\scriptsize
\renewcommand{\arraystretch}{0.92}
\setlength{\tabcolsep}{2.5pt}
\begin{tabular}{@{}p{0.25\linewidth}p{0.18\linewidth}p{0.24\linewidth}p{0.25\linewidth}@{}}
\toprule
Task & Dataset & Metric & Use in the paper \\
\midrule
Pipeline state & controlled & AUROC, CI, permutation p95 & cold/warm boundary
and cache-reuse evidence \\
Browser/native activity & controlled & macro-F1, slowdown, samples & synthetic
native activity and phase inference \\
Other WebGPU resources & controlled & AUROC, accuracy, macro-F1 & surface
coverage and negative rows \\
Participant fingerprintability & participant field study & entropy,
anonymity sets, repeats & sample-level active WebGPU distinctiveness \\
Page-load exposure & Tranco crawl & static/dynamic counts, triage &
public-page WebGPU touching \\
Mitigation & controlled and extension & AUROC reduction, latency cost &
pipeline cache policy direction \\
\bottomrule
\end{tabular}
\end{table}

\begin{table}[H]
\centering
\caption{Surface-to-impact map. Placement and observable mode determine which
claim each surface supports.}
\label{tab:surfaces-tasks}
\scriptsize
\renewcommand{\arraystretch}{0.92}
\setlength{\tabcolsep}{2.5pt}
\begin{tabular}{@{}p{0.29\linewidth}p{0.29\linewidth}p{0.29\linewidth}@{}}
\toprule
\textbf{Static exposure} & \textbf{Active contention} &
\textbf{Persistent state} \\
\midrule
features, limits, adapter metadata, graphics hashes &
bandwidth, ALU, atomics, queues, texture/frame observables &
shader and pipeline cold/warm creation state \\
\addlinespace[0.1em]
entropy and anonymity sets & AUROC, accuracy, macro-F1, slowdown &
cold/warm boundary separability \\
\bottomrule
\end{tabular}
\end{table}

\subsection{Controlled Scenario Support}
\label{app:controlled-details}
\label{app:lab-stability}

Controlled scenarios use fixed browser profiles, fixed origins, fixed window
states, selected adapter configurations, labeled victims, and explicit timer modes.
Table~\ref{tab:lab-platforms} lists the host/browser matrix used for the main
controlled claims. A row means that the browser produced valid WebGPU traces for
the listed role. Chrome is the anchor browser. Chrome and Edge carry the
Windows same-host AMD/NVIDIA comparison. Brave carries Chromium-family variant
coverage. Firefox carries browser-family coverage because Firefox/wgpu did not
expose normalized \texttt{gpu\_vendor} or \texttt{gpu\_model} fields in our
data-collection pipeline. Safari is Apple-only WebKit coverage. Multi-day repeats are
versioned in
\texttt{docs/windows-lab-stability-2026-06-11.md},
\texttt{docs/m1-lab-stability-2026-06-11.md}, and the matching
\texttt{artifact/derived/*20260611*} summaries.

\begin{table}[t!]
\centering
\caption{Controlled host/browser matrix.}
\label{tab:lab-platforms}
\scriptsize
\renewcommand{\arraystretch}{0.88}
\setlength{\tabcolsep}{2pt}
\begin{tabular}{@{}p{0.13\linewidth}p{0.12\linewidth}p{0.24\linewidth}p{0.43\linewidth}@{}}
\toprule
Host & Browser & Adapter attribution & Role in controlled matrix \\
\midrule
\multirow{4}{*}{Windows} & Chrome & AMD/RDNA-3 default; NVIDIA/Lovelace forced &
anchor Windows browser; supports same-host AMD/NVIDIA comparison across
pipeline, web-boundary, profile, browser/native, and mitigation rows \\
& Edge & AMD/RDNA-3 default; NVIDIA/Lovelace forced &
second Windows Chromium-family application for the same AMD/NVIDIA comparison;
Windows-only coverage \\
& Brave & NVIDIA/Lovelace reported by local Brave traces &
Chromium-family variant coverage: same-page, cross-tab, iframe, normal/private
CDP, and Brave$\leftrightarrow$Chrome pipeline; not used for AMD/NVIDIA
comparison \\
& Firefox & valid Firefox/wgpu traces; vendor/model not exposed &
browser-family coverage: same-page, cross-tab, iframe, and
Firefox$\leftrightarrow$Chrome pipeline; no vendor-specific AMD/NVIDIA claim \\
\midrule
\multirow{4}{*}{M1/Metal} & Chrome & Apple/Metal &
anchor Apple browser; main Apple/Metal path for surface, pipeline,
browser/native, and cross-browser rows \\
& Brave & Apple/Metal via Chromium &
Chromium-family Apple/Metal variant; Brave rows and
Brave$\leftrightarrow$Chrome pipeline \\
& Firefox & valid Firefox/wgpu traces; vendor/model not exposed &
browser-family coverage, including selected non-pipeline and
Firefox$\rightarrow$Chrome pipeline rows; no vendor-specific Apple/Metal claim
from Firefox-probe rows \\
& Safari & Safari/WebKit Apple stack &
Apple-only WebKit manual smoke row \\
\bottomrule
\end{tabular}
\end{table}

\begin{table}[t!]
\centering
\caption{Representative controlled surface rows.}
\label{tab:surface-strength}
\scriptsize
\renewcommand{\arraystretch}{0.92}
\setlength{\tabcolsep}{2.5pt}
\begin{tabular}{@{}p{0.40\linewidth}p{0.31\linewidth}r@{}}
\toprule
Surface/platform & Representative row & Metric \\
\midrule
Windows AMD pipeline & same-page high-res cold/warm & 0.986 AUROC \\
Windows NVIDIA pipeline & same-page high-res cold/warm & 0.961 AUROC \\
M1 cross-browser pipeline & aggregate high-res cold/warm & 0.718 AUROC \\
Windows NVIDIA non-pipeline & queue-progress functional unit & 0.797 AUROC \\
Windows AMD non-pipeline & queue-progress functional unit & 0.684 AUROC \\
Windows timerless non-pipeline & ordinal row & 0.575 AUROC \\
\bottomrule
\end{tabular}
\end{table}

\begin{table}[t!]
\centering
\caption{Pipeline-state controls.}
\label{tab:pipeline-controls}
\scriptsize
\renewcommand{\arraystretch}{0.92}
\setlength{\tabcolsep}{2.5pt}
\begin{tabular}{@{}p{0.30\linewidth}p{0.31\linewidth}p{0.31\linewidth}@{}}
\toprule
Control & Purpose & Coverage \\
\midrule
Mutated and unrelated shader families & separates exact reuse from generic
warmup & pipeline protocol and CI rows \\
Negative same-origin cross-tab rows & bounds boundary language & AMD and
NVIDIA near-chance rows \\
Permutation controls & checks separability under randomized labels & M1
cross-browser and native matrices \\
Multi-day repeats & checks one-session sensitivity & Windows and M1
2026-06-11 repeats \\
Multiple adapter configurations & avoids a single-adapter story & AMD/RDNA-3,
NVIDIA/Lovelace, and M1/Metal paths \\
Retaining-layer analysis & localizes the claim to browser/GPU-stack state &
browser, backend, driver, and cache layers remain separated by controls and
policy rows \\
\bottomrule
\end{tabular}
\end{table}

\paragraph*{Reading pipeline placements.}
Pipeline rows differ in what they vary. The same-page rows calibrate the
surface. Boundary rows change origin, iframe, profile, application, or browser
placement. Negative rows keep the main claim bounded.

\begin{table}[t!]
\centering
\caption{How to read pipeline placement rows.}
\label{tab:pipeline-placement-reading}
\scriptsize
\renewcommand{\arraystretch}{0.92}
\setlength{\tabcolsep}{2.5pt}
\begin{tabular}{@{}p{0.24\linewidth}p{0.30\linewidth}p{0.38\linewidth}@{}}
\toprule
Placement row & What changes & Use \\
\midrule
Same page & same origin, page, profile, and browser & surface calibration and
adapter comparison \\
Cross-origin or iframe & origin and page placement & web-boundary placement
evidence \\
Same-origin cross-tab & tab placement only in the short-interval setting &
negative control for broad boundary claims \\
Cross-profile & Chrome profile state & browser-state boundary evidence \\
Chromium-family \(A\rightarrow B\) & application wrapper with shared stack
components & small-\(n\) cache-policy confirmation \\
Apple/Metal \(A\rightarrow B\) & browser placement on the same Metal stack &
cross-browser backend check \\
\bottomrule
\end{tabular}
\end{table}

\paragraph*{Metric protocol.}
Rows use the smallest metric needed for the claim. AUROC is used for binary
cold/warm or idle/active separation. Macro-F1 is used for phase classification.
Entropy and anonymity sets are used only for participant fingerprints.

\begin{table}[t!]
\centering
\caption{Metric and uncertainty protocol.}
\label{tab:metric-protocol}
\scriptsize
\renewcommand{\arraystretch}{0.92}
\setlength{\tabcolsep}{2.5pt}
\begin{tabular}{@{}p{0.24\linewidth}p{0.30\linewidth}p{0.38\linewidth}@{}}
\toprule
Result family & Metric & Support shown \\
\midrule
Pipeline cold/warm & AUROC & cold/warm counts, 95\% CI, repeats or
permutation p95 where available \\
Browser/native activity & macro-F1, slowdown & sample counts, accuracy CI for
M1 rows, permutation maxima \\
Non-pipeline probes & AUROC or accuracy & representative positive and weak
rows, kept below pipeline in the claim hierarchy \\
Participant fingerprints & entropy, sets, singleton fraction & completed
deduplicated denominator and feature ablations \\
Mitigation rows & AUROC reduction plus cost & matched baseline, mitigated row,
latency or compatibility cost \\
\bottomrule
\end{tabular}
\end{table}

\begin{table}[t!]
\centering
\caption{Supporting browser/native throughput interference rows.}
\label{tab:browser-native-throughput}
\scriptsize
\renewcommand{\arraystretch}{0.92}
\setlength{\tabcolsep}{2.5pt}
\begin{tabular}{@{}p{0.22\linewidth}p{0.50\linewidth}r@{}}
\toprule
Platform & Supporting row & Slowdown \\
\midrule
Win NVIDIA & Native throughput; stress compute under scheduler pressure &
8.365\% \\
M1/Metal & Native throughput; stress copy under bandwidth pressure & 7.077\% \\
\bottomrule
\end{tabular}
\end{table}

\subsection{Participant and Crawl Support}
\label{app:participant-ingestion}
\label{app:crawl-details}

The participant field study contributes deployment, compatibility,
sample-level fingerprintability, and repeat evidence. The export contains
7,999 participant-upload rows, 1,534 reconstructed logical runs, 1,127 completed
runs, 1,095 completed deduplicated records, and 80 local-storage identifiers
seen more than once. The local-storage identifier describes the same browser
storage state returning to the benchmark. It is separate from account, IP
address, browser product, and physical-device identity. Signature construction
excludes participant identifiers, upload filenames, chunk counts, form
metadata, timestamps, completion code, participant-platform metadata, and
recruitment URL parameters.

\begin{table}[t!]
\centering
\caption{Participant funnel and crawl interpretation rules.}
\label{tab:field-funnel}
\label{tab:crawl-detection-levels}
\label{tab:crawl-triage-rules}
\scriptsize
\renewcommand{\arraystretch}{0.92}
\setlength{\tabcolsep}{2.5pt}
\begin{tabular}{@{}p{0.36\linewidth}p{0.56\linewidth}@{}}
\toprule
Item & Interpretation \\
\midrule
1,095 completed deduplicated records & denominator for field fingerprintability
and entropy rows \\
80 repeated local-storage identifiers & repeat evidence for the same local
browser storage state \\
Static crawl reference & WebGPU terms in fetched scripts; bundled support,
feature detection, or dormant logic \\
Adapter/device probe & runtime \texttt{requestAdapter} or
\texttt{requestDevice}; environment signal, not submitted work \\
Observed work & shader, pipeline, queue, query, or map activity; count is zero
in the page-load crawl \\
Source-context labels & fingerprinting, anti-abuse, media/player, analytics,
or unclear context; source context only \\
\bottomrule
\end{tabular}
\end{table}

\paragraph*{Repeat and signature semantics.}
The participant export uses a first-party local-storage identifier for repeat
control. A repeated identifier means the same browser storage state returned to
the benchmark. Fingerprint rows use analysis objects built after upload and
recruitment fields are removed.

\begin{table}[t!]
\centering
\caption{Participant repeat and signature semantics.}
\label{tab:participant-repeat-semantics}
\scriptsize
\renewcommand{\arraystretch}{0.92}
\setlength{\tabcolsep}{2.5pt}
\begin{tabular}{@{}p{0.28\linewidth}p{0.31\linewidth}p{0.33\linewidth}@{}}
\toprule
Object & Meaning & Use \\
\midrule
Participant-upload row & one uploaded chunk or compact payload row & raw input
to logical-run reconstruction \\
Logical run & reconstructed browser benchmark execution & completion and
compatibility accounting \\
Deduplicated completed record & one completed record per local-storage group
for primary field metrics & fingerprintability denominator \\
Repeated local-storage identifier & same first-party storage state seen again &
repeat stability check \\
Output-only WebGPU signature & render, compute, and pipeline output hashes &
deterministic active components \\
Active WebGPU signature & outputs, stability flags, and compact timing
summaries & sample-level active distinctiveness \\
\bottomrule
\end{tabular}
\end{table}

The crawl records four evidence levels. Static references show that crawled
code contains WebGPU terms. Adapter and device probes show runtime API use.
Observed work requires shader, pipeline, queue, query, or mapping activity.

\begin{table}[t!]
\centering
\caption{Crawl hook coverage.}
\label{tab:crawl-hook-coverage}
\scriptsize
\renewcommand{\arraystretch}{0.92}
\setlength{\tabcolsep}{2.5pt}
\begin{tabular}{@{}p{0.30\linewidth}p{0.31\linewidth}p{0.31\linewidth}@{}}
\toprule
Observed surface & Hook or scan point & Interpretation \\
\midrule
Static WebGPU terms & fetched script scan & bundled support, detection, or
dormant code \\
\texttt{navigator.gpu} access & API presence hook & WebGPU considered by page
code \\
\texttt{requestAdapter} & adapter request hook & capability or environment
probe \\
\texttt{requestDevice} & device request hook & stronger capability probe \\
Shader/pipeline work & shader-module and pipeline hooks & page-load WebGPU work
when present \\
Queue/query/map work & queue, query-set, and buffer-map hooks & submitted or
observable GPU work when present \\
\bottomrule
\end{tabular}
\end{table}

The crawl attempted 10,000 Tranco URLs, obtained 7,477 successful page-load
records, and found 56 WebGPU-positive records: 33 with dynamic WebGPU calls and
29 with static evidence. None showed page-load shader-module creation, pipeline
creation, queue submission, query-set use, or buffer mapping. A focused
reproduction pass on 13 representative positives reproduced runtime WebGPU
calls on 12 targets; Chrome, Edge, and Brave preserved the same outcome and
call count for all 13 targets.

\subsection{Mitigation Mechanics}
\label{app:mitigation-supplement}

Mitigation rows are policy simulations and extension-level demonstrations. The
Windows source-level key-separation row reduces AUROC from 0.994 to 0.606 at
+6.05 ms warm-compile cost. The stronger M1 condition-level row reduces AUROC
from 0.759 to 0.520 at +0.53 ms warm-compile cost. The Chromium extension
source-partition prototype reduces AUROC from 0.648 to 0.525. These rows point
to pipeline cache partitioning as the clearest measured direction; timer
clamping and dispatch caps remain exploratory.

\begin{table}[t!]
\centering
\caption{Defense families and browser targets. The extension implements these
as demonstrations; production versions belong inside the browser, backend, or
API policy layer.}
\label{tab:defense-families}
\scriptsize
\setlength{\tabcolsep}{3pt}
\begin{tabular}{@{}p{0.17\linewidth}p{0.20\linewidth}p{0.33\linewidth}p{0.22\linewidth}@{}}
\toprule
Defense family & Target surface & Browser-level analogue & Functionality cost \\
\midrule
Source partitioning & pipeline cold/warm state & origin/profile/private-mode
pipeline-cache partitioning or cache-key separation & compile latency; less cache
reuse \\
Metadata bucketing & static adapter exposure & feature, limit, adapter, and
format bucketing & lower capability precision; possible compatibility fallback \\
Observable padding & queue, map, and completion timing & delayed or bucketed
completion semantics & latency, frame pacing, and jank risk \\
Readback perturbation & active render/compute fingerprints & explicit privacy
mode or readback budget for fingerprint-sensitive surfaces & correctness risk
for applications that expect exact readback \\
Access gating & all WebGPU surfaces & permission, visibility, activation, or
enterprise policy gate & highest compatibility loss; blocks legitimate WebGPU \\
\bottomrule
\end{tabular}
\end{table}

\begin{table}[t!]
\centering
\caption{Exploratory timer and dispatch mitigation rows.}
\label{tab:mitigation-exploratory}
\scriptsize
\renewcommand{\arraystretch}{0.92}
\setlength{\tabcolsep}{2.5pt}
\begin{tabular}{@{}p{0.18\linewidth}p{0.29\linewidth}p{0.25\linewidth}p{0.20\linewidth}@{}}
\toprule
Surface & Policy or condition & AUROC base$\rightarrow$mit. (delta) & Cost \\
\midrule
Scheduler & 32 ms timer clamp pilot & 0.637$\rightarrow$0.500 (0.137) &
32 ms precision \\
Bandwidth & 8 ms timer clamp pilot & 0.588$\rightarrow$0.644 (-0.056) &
8 ms precision \\
Scheduler & dispatch-cap pilot & 0.506$\rightarrow$0.546 (-0.039) &
cap machinery only \\
\bottomrule
\end{tabular}
\end{table}

\begin{table}[t!]
\centering
\caption{Extension-level mitigation mechanics in WGPULens Shield.}
\label{tab:extension-mechanics}
\scriptsize
\renewcommand{\arraystretch}{0.92}
\setlength{\tabcolsep}{2.5pt}
\begin{tabular}{@{}p{0.25\linewidth}p{0.31\linewidth}p{0.36\linewidth}@{}}
\toprule
Defense & Patched WebGPU surface & Purpose and tradeoff \\
\midrule
Pipeline partition & shader-module creation & appends a valid module-scope
constant derived from \texttt{location.origin}; reduces cross-origin
shader/pipeline reuse and may add compilation work \\
Static masking & adapter request/info proxy & buckets adapter metadata,
features, and limits; active probes remain possible \\
Observable padding & async pipeline, queue, and map promises & resolves
promises on coarse 32 ms buckets; preserves results but adds latency \\
Readback noise & mapped buffer readback & perturbs sparse mapped bytes to
disrupt active output hashes; can affect exact compute correctness \\
WebGPU gate & adapter request & returns \texttt{null}; strongest extension
option and largest compatibility cost \\
\bottomrule
\end{tabular}
\end{table}

\paragraph*{Extension operating modes.}
The extension exposes each defense as a check. Presets combine them into three
practical modes: balanced keeps outputs exact, strict perturbs readback, and
gate removes WebGPU from the page.

\begin{table}[t!]
\centering
\caption{WGPULens Shield operating modes.}
\label{tab:extension-modes}
\begin{minipage}{0.96\linewidth}
\centering
\scriptsize
\renewcommand{\arraystretch}{0.84}
\setlength{\tabcolsep}{2.5pt}
\begin{tabular}{@{}p{0.23\linewidth}p{0.38\linewidth}p{0.29\linewidth}@{}}
\toprule
Mode & Enabled defenses & Compatibility profile \\
\midrule
Observe & counters and page status & demonstration and comparison \\
Balanced & pipeline partition, static mask, observable padding & exact
render/compute outputs; added latency and metadata reduction \\
Strict anti-fingerprinting & balanced mode plus readback noise & active-output hashes
change; exact readback can change \\
Gate & adapter request unavailable & strongest reduction; highest functionality
cost \\
\bottomrule
\end{tabular}
\end{minipage}
\end{table}

\paragraph*{Partition granularity.}
The pipeline measurements separate reuse within a privacy boundary from reuse
across one. Browser implementations should treat shader and pipeline reuse as a
privacy-policy decision, not only as a compiler optimization. At minimum, the
cache key should include the selected top-level site or origin partition,
profile, private-mode/session state, adapter/backend identity, and pipeline
descriptor material. Same-boundary reuse preserves most cache benefit.
Cross-boundary reuse exposes the persistent-state signal measured in
Table~\ref{tab:pipeline-boundary}. Timer coarsening and readback policy address
different observables; they do not replace cache-key separation for persistent
pipeline state.

The shim runs as a main-world content script and installs once WebGPU types are
available. Its direct coverage is page code after injection. Worker coverage,
native applications, and OS-level GPU state require browser-engine support. A
browser-engine version would move partitioning into shader/pipeline cache keys
and expose metadata, timing, and readback policy through browser privacy
controls.

\begin{table}[t!]
\centering
\caption{Browser-engine mitigation targets.}
\label{tab:browser-mitigation-targets}
\begin{minipage}{0.96\linewidth}
\centering
\scriptsize
\renewcommand{\arraystretch}{0.92}
\setlength{\tabcolsep}{2.5pt}
\begin{tabular}{@{}p{0.31\linewidth}p{0.28\linewidth}p{0.33\linewidth}@{}}
\toprule
Browser target & State or observable & Engineering decision \\
\midrule
Cache-key partitioning & shader/pipeline caches & include origin, profile,
session, and private-mode policy state in reuse decisions \\
Metadata reduction & adapter info, features, limits & bucket or budget precise
capabilities while preserving common compatibility tiers \\
Async observable policy & pipeline, queue, map timing & coarsen, budget, or
normalize completion observables by privacy mode \\
Readback policy & mapped GPU results & keep exact by default; use strict
anti-fingerprint behavior with compatibility signaling \\
Access policy & adapter availability & use permission, visibility, privacy
budget, or enterprise policy when WebGPU access is unavailable \\
\bottomrule
\end{tabular}
\end{minipage}
\end{table}

\end{document}